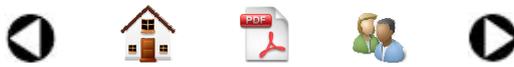

# Dynamic OFDMA Resource Allocation for QoS Guarantee and System Optimization of Best Effort and Non Real-time Traffic

Arijit Ukil[1], Jaydip Sen[2], Debasish Bera[3]

Wireless and Multimedia Innovation Lab, Tata Consultancy Services

Kolkata, India

Email: {arijit.ukil[1], jaydip.sen[2], debasish.bera[3] @tcs.com}

*Abstract*— **To achieve the requirement of high data rate, low latency, user fairness for next generation wireless networks, proper designing of cross-layer optimized dynamic resource allocation algorithm is prerequisite. In this paper, we propose a dynamic resource allocation scheme in Orthogonal Frequency Division Multiple Access (OFDMA) systems to optimize the non real-time (NRT) traffic, which requires allocation of minimum quantum of data within a predefined time that does not incur packet loss. Most existing and proposed works on resource allocation schemes focused on traffic consisting of delay-constraint real-time (RT) or delay-tolerant (NRT, Best-Effort (BE)) applications in a single scheme. In this work, we investigate the resource allocation problem in heterogeneous multiuser OFDMA system with the objective of optimizing the aggregate data delivery of NRT and BE traffic to maximize the overall system performance, by exploiting the inherent time-diversity gain in mobile wireless environment for delay-tolerant applications. Simulation results show that the proposed algorithm greatly enhances the system capacity, when compared to traditional proportional fair resource allocation algorithm.**

## I. INTRODUCTION

Rapid growth of wireless technology, coupled with the explosive growth of the Internet, has increased the demand for wireless data services. Traffic on beyond 3G wireless networks is heterogeneous with random mix of RT and NRT traffic with applications require widely varying and diverse quality of service (QoS) guarantees for the different types of offered traffic, which enforces a robust and application specific optimization of limited system resources. For wireless QoS guarantee, link or channel adaptation techniques along with dynamic resource allocation have been widely considered as the key solution to overcome the impact of the unpredictability of the wireless channel. Scheduling or resource allocation plays an important role in providing QoS support in various kinds of wireless networks [1]. Various scheduling disciplines have been developed in order to guarantee certain required QoS over wired networks, which are not suitable in wireless environment due to its time-varying channel capacity.

Orthogonal Frequency Division Multiple Access (OFDMA) is the de facto standard multiple access scheme for next generation wireless standards like WiMAX, LTE, IMT-A. OFDMA, also referred to as multi-user OFDM is normally characterized by a fixed number of orthogonal subcarriers to be allocated to the available users [2]. Resource allocation algorithms in OFDMA dynamically assign radio resources to the users from apriori knowledge of the channel condition according to the system objective function. Dynamic resource allocation is a kind of cross layer optimization mainly involving Physical (PHY) and Media Access Control (MAC) manages the system resources, like bandwidth, transmit power by exploiting the frequency and temporal dimension of the resource space adaptively to achieve the system performance objective. In [3], theoretical frame-work and development of effective algorithms for efficient and fair resource allocation in OFDM wireless networks under different utility conditions has been discussed. To maximize the system's overall performance, most of the radio resources have to be allocated to the users with good channel condition, which in turn arise fairness issue [4]. So, a trade-off is required between system performance and fairness among the users. In literature, this trade-off is tackled by the concept of Proportional Fair (PF) [5]. Another important factor is QoS provisioning. QoS-aware fair resource allocation in OFDMA systems has been proposed in [6, 7]. So, we find that at least three degrees of freedom exist to properly optimize a wireless system, viz. performance maximization, fairness and QoS provisioning. Most of the present authors proposed resource allocation algorithm for traffic model containing a mixture of RT and NRT applications and attempted to optimize the system performance by scheduling them at the same time [10] through a single independent scheme. But QoS requirement for these two classes are very dissimilar. RT traffic is characterized by constant bandwidth allocation and hard delay-bound, where as in NRT traffic, mostly minimum bandwidth is to be allocated with no stringent delay-bound. So, it is obvious that the resource allocation algorithm should depend on the traffic classes as a delay-differentiated algorithm.

In this paper, we present an OFDMA resource allocation scheme specifically dedicated to optimize the traffic classes with less delay-sensitivity. Instead of instantaneous QoS assurance we propose to guarantee the QoS requirement in an aggregate sense, for an example, a file of certain size ( X kb) has to be transferred within a certain time limit (Y sec), but at any instant X/Y kbps is not guaranteed, but the file is guaranteed to be delivered within Y sec. Here, we exploited the time-diversity gain, an inherent characteristics of highly dynamic wireless mobile channel, which has very less coherence time. We present an associated resource allocation algorithm and compare the proposed algorithm with popular PF algorithm from the literature. We observe that our proposed algorithm optimizes the system better and produces better performance gain. The paper is organized as follows. The next section describes the system model. In section III traditional PF optimization is discussed. Aggregate throughput optimization (ATO) and algorithm are presented in detail in section IV. Simulation results and analysis of the ATO algorithm are presented in Section V. Section VI provides summary and conclusion.

## II.  SYSTEM MODEL

The considered OFDMA system consists of single cell with one base station (BS) communicating simultaneously with $K$ user terminals using N OFDMA subcarriers, shown in Fig. 1.

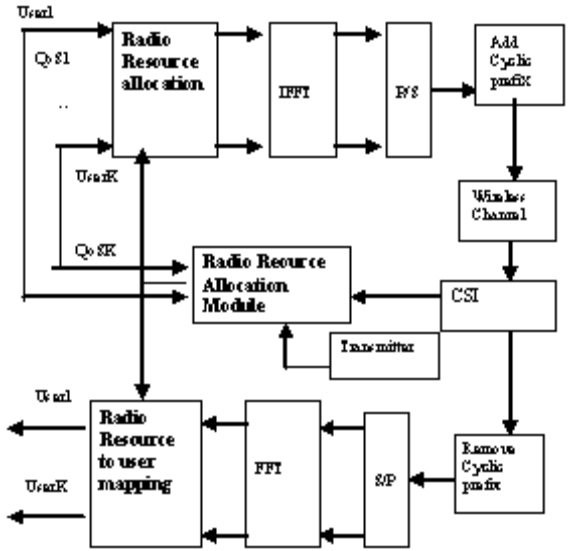

Figure 1.  Multi-user OFDMA System Model

The interference from adjacent cells is treated as background noise. Total noise power density including background noise and AWGN noise is taken as $N_T$. The mutually disjoint OFDMA subcarriers are denoted as: $\Omega_1$, $\Omega_2$, ....$\Omega_N$ , where $\Omega_n = $ B/N and $\Omega_n \leq$ Bc , where Bc is the coherence bandwidth of the channel and B is the total available bandwidth. $P_T$ is the total available transmit power and $P_{kn}$ is the transmit power for $n^{th}$ subcarrier when transmitted to $k^{th}$ user, where $P_{kn} = $ $P_T$/N, as performance can hardly be deteriorated by equal power distribution [7]. Perfect channel characteristic is assumed in the form of channel state information (CSI). Channel Gain for subcarrier n for user k at $t^{th}$ allocation instant is taken as $h_{knt}$, which is estimated from CSI information. Let $\omega_{kt}$ be the achievable rate for $k^{th}$ user at $t^{th}$ instant, then $\omega_{kt}$ is a function of the channel condition or $h_{knt}$ and can be expressed as:

$$\omega_{kt} = \sum_{n=1}^{N} \Omega_n \times \rho_{knt} \times f(h_{knt}) \tag{1}$$

where $f(h_{knt})$ can be expressed as below and $\rho_{knt}$ is the sub-carrier assignment matrix, which is equal to 1 if $n^{th}$ subcarrier assigned to $k^{th}$ user at $t^{th}$ time instant, else equal to 0.

$$\sum_{k=1}^{K} \rho_{knt} = 1 \text{ and } \sum_{n=1}^{N} \sum_{k=1}^{K} \rho_{knt} = N$$

$$f(h_{knt}) = \log_2 \left( 1 + \frac{h_{knt}^2 \times P_{kn} \times SNR\_gap}{N_T \times \Omega_n} \right)$$

In practice the achieved data rate is less than that of what equation (1) suggests as there exists a few dB SNR gap. SNR_gap is the imperfection of theoretical value of achievable data rate to the actual data rate. SNR_gap in simplified [3] term can be approximated as $\frac{-\ln(5BER)}{1.6}$. The QoS parameter for BE or NRT traffic of $k^{th}$ user is defined as $[\gamma_k, \Delta_k]$ , where $\gamma_k, \Delta_k$ are the minimum total throughput requirement and data-transfer-duration for user k. ATO is performed in long-term; which is $\Delta_{ato} = $ Min $(\Delta_k), \forall k$ , which is basically consisting of a number of allocation epochs (=frame-duration) and $\Delta_{ato} = M \times T_f$ , M = 1, 2 ..., where $T_f$ is the frame-duration, which is taken as the unit allocation duration. $T_f \leq \Gamma_c$ , where $\Gamma_c$ is the coherence time of the channel. The objective of ATO is to optimize the aggregate data size $\omega_k$, delivered to individual user over $\Delta_{ato}$ time duration subject to the user's QoS constraint: 
$$\omega_k = \sum_{t=0}^{\Delta_{ato}} \omega_{kt} \tag{2}$$

## III.  PROPORTIONAL FAIR OPTIMIZATION FOR QoS PROVISIONING

The advantages of multiuser diversity only contribute to a small portion of mobile users whose channel quality is good, which may not lead to a significant QoS performance improvement from entire network perspective. Proportional fair (PF) is considered to be a simple yet effective fairness notion. PF optimization is a pure outcome fairness metric, which is simple to use, but does not guarantee fairness in a strict sense. PF utility is defined as a logarithmic function of the rate allocated to a user. Because of the convex nature of the logarithmic function, diminishing return is modeled. Proportional Fairness Index (PFI) is a kind of optimized measure of user fairness. PFI at $t^{th}$ allocation instant for $k^{th}$ user is expressed as [5]:

$$PFI_{kt} = \frac{\omega_{kt}}{\overline{\omega}_{kt}}$$

The objective of PF optimization is to allocate subcarrier n to k* user when [9]:

$$k* = \arg\max_k \frac{\omega_{kt}}{\overline{\omega}_{kt}} \qquad (3)$$

$$\overline{\omega}_{kt} = (1 - \frac{1}{\Delta \tau})\overline{\omega}_{k(t-1)} + \frac{1}{\Delta \tau}\omega_{k(t-1)}$$

The mean data rate achieved ($\overline{\omega}_{kt}$) is computed as moving average with the purpose of providing fairness [9]. System constraint to provide QoS is: $\omega_{kt} \geq \gamma_k$, $\forall k$.

PF optimization described above, fairly distributes radio resources to guarantee instantaneous QoS without any differentiation based on user traffic-class. PF optimization is fundamentally a generalized allocation scheme, independent of user traffic characteristics, so not suitable for heterogeneous traffic.

## IV. AGGREGATE THROUGHPUT OPTIMIZATION AND ALGORITHM FOR NRT AND BE TRAFFIC

The objective of aggregate throughput optimization is to maximize the overall system performance by optimizing the radio resource requirement of delay-tolerant traffic classes. Having very different characteristics, each application deserves a specific degree of service, defined at the application layer. Several standardization bodies have tried to define service categories. ETSI Project TIPHON [ETSI-TR102] proposes an alternative QoS class definition, shown in Table 1. Statistical QoS guarantee based adaptive resource allocation has already been proposed [8] as an attempt to efficiently support the diverse QoS requirements with considerable performance improvement, where the complexity increases linearly with KN. The proposed scheme in [8] considers only the real-time traffic to guarantee the statistical delay-bound for heterogeneous mobile users, but the last two classes of traffic are not given due importance.

TABLE I
TIPHON QoS classes (from [ETSI-TR102])

| QoS class | Components | QoS characteristics |
|---|---|---|
| Real-time conversational (telephony, Audio and video conference) | Speech, audio, video, multimedia | Delay and delay variation sensitive, limited tolerance to loss and errors, constant and variable bit rate. |
| Real-time streaming(audio and video broadcast, surveillance) | Audio, video, multimedia | Tolerant to delay, delay variation sensitive, variable bit rate |
| Near real-time interactive ( web browsing) | Data | Delay sensitive, tolerant to delay variation, error sensitive, variable bit rate |
| Non-real-time background (e-mail, file transfer) | Data | Not delay and delay variation sensitive, error sensitive, best effort |

Here, we introduce the concept of aggregate throughput guarantee for relaxed delay-bound traffic classes. As the relaxed or soft delay-bound is the property of NRT and BE classes, so our scheme will optimize NRT and BE traffic and release most of the radio resources to the demanding traffic classes, like Unsolicited Grant Services (UGS) and RT. By radio resource, we will mean OFDMA subcarriers. The scheme intelligently assigns mutually disjoint subcarriers to the users from apriori knowledge of the channel condition by taking the advantage of multi user diversity and maximizes the overall system optimization by exploiting time-diversity gain of fast-fading wireless mobile environment. Basically by principal, a well designed communication system should take the available degrees of freedom of the channel as much as possible. Wireless channel is normally very much dynamic in nature and over long duration, time diversity gain becomes high and the mean channel condition ($\overline{h}_{kn}$) follows similar distribution according to the Bernoulli's Law of Large Numbers. As $h_{kn}$ is commonly considered as independent and identically distributed (iid) random variable with mean $\phi_k$, then theoretically, $\underset{\Delta_{aoa} \to \infty}{Lim} TDgain = \phi_k$. Time diversity is achieved by averaging the fading of the channel over time and it can be proved that time-diversity incurs substantial gain in the detection of the signal or in other words, improves probability of error significantly. Considering a flat fading channel, L time-diversity branches, $x_l$ as the transmitted signal at $l^{th}$ time diversity branch, $h_l$ as the channel gain and $w_l$ as the additive noise, $y_l$ the received signal is:

$$y_l = h_l \times x_l + w_l$$

where, $w_l$ as iid random variable, $N(0, N_T)$.

Probability of error of detecting $x_l$ is given by:

$$P_e(x_l) = Q\left(\sqrt{2x_l SNR}\right)$$

where Q(.) is complementary cumulative distribution function of a $N(0, N_T)$ random variable. If $\|h\|^2 \times SNR$ is the received SNR for a given channel vector $\mathbf{h}$, under Rayleigh fading with each gain $h_l$ is iid N (0, 1) random variable, then:

$$P_e(x_l) = Q\left(\sqrt{2\|h\|^2 SNR}\right)$$

where, $\|h\|^2 = \sum_{l=1}^{L} |h_l|^2$.

It can be proved that theoretical achievable data rate increases linearly with the number of diversity branches. Let $\omega_k^L, \omega_k^1$ be the user achievable data rate in the presence of L and 1 number of diversity branches respectively. Then at SNR >> 1,

$$\omega_k^L \approx \left(L \times \omega_k^1\right) \qquad (4)$$

It can also be proved that for a fixed SNR:

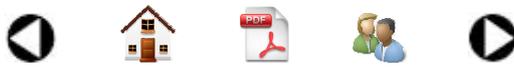

$$\overline{P_e} = \frac{\kappa'}{\left(2^{\frac{\overline{\omega_k^i}}{\overline{\xi} \times \Omega_n \times L}}\right)^L} \qquad (5)$$

where $\xi, \kappa'$ are constants.

From (4), it can be observed that:

$$\underset{L \to \infty}{Lim} \overline{P_e(x_l)} = 0$$

The proof can not be given in detail due to limitation of space. Replacing L by M as time-diversity branches, where M is the number of allocation instants to be considered for resource allocation to satisfy the aggregate throughput. It can be noted that more the value of M, the ergodicity increases. Time diversity is a less studied diversity technique as it needs delay-tolerant applications and highly dynamic mobile wireless environment for some significant performance gain. One of its main advantages when compared to other diversity techniques is that it does not require additional equipment or complicated synchronization procedures, since it involves only a single channel and a single reception path. From (4,5), it can be concluded that arbitrary large data rate can be achieved as well less probability of error can be reached with large value of M. But M introduces latency in the system, which is a limiting factor for exploiting the advantage of obtaining time diversity gain.

From this background, our proposed optimized scheme is:

$$\max \sum_{n=1}^{N} \sum_{k=1}^{K} \frac{\omega_{kn}}{\sum_{t=0}^{\Delta_{ato}} \omega_{knt}} \qquad (6)$$

$$\omega_k = \sum_{t=0}^{\Delta_{ato}} \sum_{n=1}^{N} \rho_{knt} \times \omega_{knt} \qquad (7)$$

$$\sum_{t=0}^{\Delta_{ato}} \sum_{n=1}^{N} \sum_{k=1}^{K} \rho_{knt} = N \times M \qquad (8)$$

subject to:

$$\omega_k \geq \gamma_k \ , \ \forall k \qquad (9)$$

$$\Delta_{ato} \gg \Gamma_c, \Delta_{ato} \leq \max tolerable(\Delta_k), \forall k \qquad (10)$$

Equation (6) is the objective of the optimization scheme, which states that 'maximize the aggregate data achievable to individual user in every allocation epoch with a notion of proportional fairness (3). Equation (8) is to monitor mutual exclusiveness of sub-carrier allocation to the users. The optimization scheme has to meet the QoS constraints of the users (9-10). Equation (9) is to examine that user's minimum data requirement is properly delivered within the long-term duration, $\Delta_{ato}$ where as (10) is required for availing time-diversity gain subject to maximum tolerable delay limit of all the users are satisfied. Equation (10) is the limitation of the optimization scheme. System should estimate a threshold for

$\Delta_{ato}$ from the long-term channel statistics and will not allow users with aggregate data delivery delay-bound limit below that threshold. It can also be observed from (5) that more time diversity gain would be achieved with the increasing value of M, i.e. $\underset{\Delta_{ato} \to \infty}{Lim} P\left(\frac{\omega_k}{\gamma_k}\right) = 1$. So, we confirm that the value of $\Delta_{ato}$ is the limiting factor to extract the best out of ATO. Systems with very much constraint (less) value of $\Delta_{ato}$ will be much less optimized than a system with high value of $\Delta_{ato}$, but high level of channel dynamics may compensate the performance loss.

From the optimization scheme proposed above (6-10), the corresponding resource allocation algorithm ATO is:

1. *Initialize* $\omega_{knt} = \psi$, $\omega_{kt} = 0$, $\forall k$, *where* $\psi$ *is a random number*.

2. *Assign initial subcarriers to the users at t = 0.*

$$while \ (\{\Omega_n\} \neq \Phi \ ) \ do$$
$$for \ k= 1: K$$
$$calculate \ \ \omega_{knt} = f(h_{knt})$$
$$end \ for$$
$$k* = \arg\max_k(\omega_{knt})$$
$$\omega_{k*t} = \omega_{k*nt} + \omega_{k*t}$$
$$end \ while$$

3. *Aggregate data delivery in* $\Delta_{ato}$

$$for \ t = T_f : \Delta_{ato}$$
$$while \ (\{\Omega_n\} \neq \Phi \ ) \ do$$
$$for \ k= 1: K$$
$$if \ ( \ \omega_k < \gamma_k \ )$$
$$k* = \arg\max_{k,n} \frac{\omega_{kn}}{\sum_{t=T_f}^{\Delta_{ato}} \omega_{knt}}$$
$$end \ for$$
$$\omega_{k*t} = \omega_{k*nt} + \omega_{k*t}$$
$$end \ while$$
$$calculate \ \omega_k \ as \ per \ (7)$$
$$end \ for$$

## V.  SIMULATION AND ANALYSIS

In this section we present simulation results of the proposed ATO algorithm in MATLAB environment under the system parameters given in Table 2. The system parameters are roughly based on Mobile WiMAX Scalable OFDMA-PHY. Performance comparison is studied by increasing the data-delivery duration, which in turn increases the total data delivery size, i.e., when $\Delta_k \to \mu \times \Delta_k$, $\gamma_k \to \mu \times \gamma_k$.

TABLE II
Simulation and System Parameters

| Available Bandwidth | 1.25 MHz |
|---|---|
| Total Transmitted Power | 20 dBm |
| Number of users | 20 |
| Number of sub-carriers | 72 |
| BER | $10^{-3}$ |
| Frame duration | 5 msec |
| Channel model | Rayleigh |
| Modulation | 16QAM |
| Channel sampling frequency | 1.5 MHz |
| Maximum Doppler | 100Hz |

Fig. 2 depicts the plot comparing the total data delivered to the users when two allocation instants are considered, i.e., M=2. Here it is apparent that the achieved total data deviates from the QoS profile considerably for both general PF (GPF) and ATO algorithm. Fig. 3 shows the result when 20 allocation instants are considered. It is clear that here performance improvement for ATO algorithm is much more than GPF, and ATO follows the QoS parameter very closely. ATO attempts to converge to the total data delivery value $\gamma_k$ with the increase of time-length by exploiting time diversity gain, where as GPF is virtually independent of time-length ($\Delta_{ato}$).

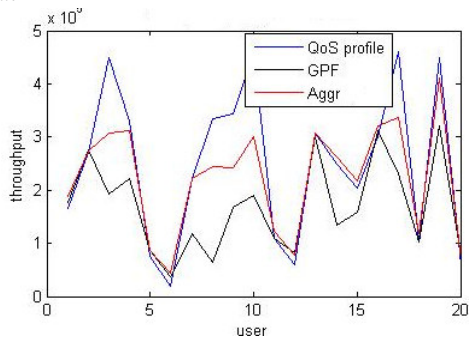

Figure 2. Performance comparison of ATO and GPF algorithm at M=2

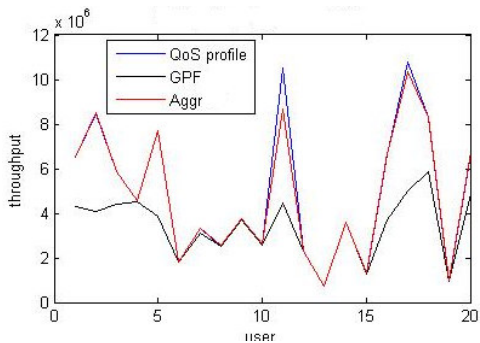

Figure 3. Performance comparison of ATO and GPF algorithm at M =20

Fig. 3 establishes the claim of better performance by ATO algorithm compared to GPF. It has to be noted that for RT traffic with hard delay-bound, i.e., when M=1, ATO does not improve much over GPF. With increasing value of M, the convergence of aggregate data delivered to QoS parameter will become closer and significant difference of performance achievement compared to GPF will be more prominent. Fig 4 is the corresponding CDF (Cumulative Distribution Function) plot.

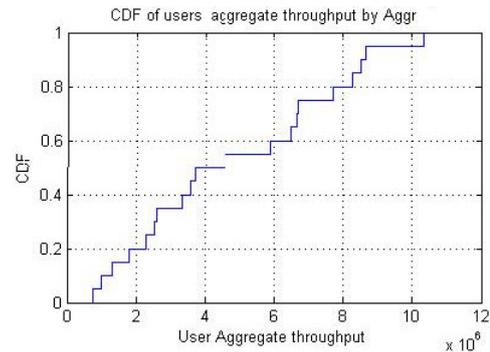

Figure 4. CDF plot of aggregate data delivery to the users at M=20

## VI. SUMMARY, CONCLUSION AND FUTURE SCOPE OF WORK

Here an efficient and optimized OFDMA resource allocation algorithm is presented, which has shown the characteristics of better performance for NRT and BE traffic when compared to GPF algorithm in QoS diversified heterogeneous traffic condition. The only limitation is the maximum bound of the long term duration. ATO algorithm is simple, low complexity and of very practical importance and can be implemented for next generation broadband wireless systems like LTE, WiMAX, IMT-A. Future scope of work lies in implementing this algorithm to a next generation wireless system and evaluating the performance through real-life field trials.

## REFERENCES

[1] Qingwen Liu, Xin Wang, and Georgios B. Giannakis, "A Cross-Layer Scheduling Algorithm With QoS Support in Wireless Networks", IEEE Trans. in Vehicular Technology, vol. 55, no. 3, pp. 839-847, May 2005.

[2] K. David Astély et al., "A Future Radio-Access Framework", IEEE JSAC, vol. 24, no. 3, pp. 693-706, March 2006.

[3] Guocong Song,Ye (Geoffrey) Li,"Cross-Layer Optimization for OFDM Wireless Networks—Part II: Algorithm Development", IEEE Trans. on Wireless Comm., vol. 4, no. 2, pp. 625-634, March 2005

[4] W. Rhee and J. M. Cioffi. "Increase in capacity of multiuser OFDM system using dynamic subchannel allocation". Proc., IEEE VTC 2000, Page: 1085–89.

[5] Tien-Dzung Nguyen and Youngnam Han."A Proportional Fairness Algorithm with QoS Provision in Downlink OFDMA Systems". IEEE Communication Letters. Vol-2, No.-11, Nov 2006.

[6] Mustafa Ergen, Sinem Coleri, and Pravin Varaiya, "QoS Aware Adaptive Resource Allocation Techniques for Fair Scheduling in OFDMA Based Broadband Wireless Access Systems", IEEE Trans. On Broadcasting, vol. 49, no. 4, pp. 362-370, Dec 2003

[7] Wei Xu, Chunming Zhao, Peng Zhou, and Yijin Yang ."Efficient Adaptive Resource Allocation for Multiuser OFDM Systems with Minimum Rate Constraints"., ICC, pp. 5126-5131, 2007.

[8] Jia Tang, Xi Zhang, "Cross-Layer-Model Based Adaptive Resource Allocation for Statistical QoS Guarantees in Mobile Wireless Networks", IEEE Trans. on Wireless Comm., vol. 6, no. 12, Dec 2007.

[9] Christian Wengerter, Jan Ohlhorst, Alexander Golitschek Edler von Elbwart., "Fairness and Throughput Analysis for Generalized Proportional Fair Frequency Scheduling in OFDMA.", IEEE VTC, 2005,Vol.3,pp.1903-1907.

[10] Lei Huang et al. , "Adaptive Resource Allocation for Multimedia QoS Management in Wireless Networks", IEEE Trans. on Vehicular Technology, vol. 53, no. 2, pp: 447-458, March 2004